# Enhancing Adaptive Video Streaming through Fuzzy Logic-Based Content Recommendation Systems: A Comprehensive Review and Future Directions

Koffka Khan[1]

[1]Department of Computing and Information Technology, Faculty of Science and Agriculture, The University of the West Indies, St. Augustine Campus, TRINIDAD AND TOBAGO.
Email address: koffka.khan@gmail.com

*Abstract*— As the demand for high-quality video content continues to rise, adaptive video streaming plays a pivotal role in delivering an optimal viewing experience. However, traditional content recommendation systems face challenges in dynamically adapting to users' preferences, content features, and contextual information. This review paper explores the integration of fuzzy logic into content recommendation systems for adaptive video streaming. Fuzzy logic, known for handling uncertainty and imprecision, provides a promising framework for modeling and accommodating the dynamic nature of user preferences and contextual factors. The paper discusses the evolution of adaptive video streaming, reviews traditional content recommendation algorithms, and introduces fuzzy logic as a solution to enhance the adaptability of these systems. Through a comprehensive exploration of case studies and applications, the effectiveness of fuzzy logic in improving user satisfaction and system performance is highlighted. The review also addresses challenges associated with the integration of fuzzy logic and suggests future research directions to further advance this approach. The proposed framework offers insights into a dynamic and context-aware content recommendation system, contributing to the evolution of adaptive video streaming technologies.

## I. INTRODUCTION

Adaptive video streaming is a dynamic content delivery mechanism designed to provide users with an optimal viewing experience by adjusting the quality of video content based on changing network conditions, device capabilities, and user preferences. Unlike traditional streaming methods, adaptive streaming allows for real-time adjustments, ensuring smoother playback and minimizing buffering disruptions. This technology has become increasingly crucial in the era of online multimedia consumption, where users expect seamless access to high-quality content across a variety of devices and network environments.

While adaptive video streaming significantly enhances the user experience, content recommendation [33], [12], [27], [13], [25] in this context faces multifaceted challenges. User preferences, content features, and contextual factors are dynamic and can vary widely, making it challenging to predict and respond effectively to viewers' needs. Traditional recommendation algorithms often struggle to adapt quickly to changes in user behavior or shifts in content popularity. Additionally, the diversity of available content and the need for real-time adjustments pose significant hurdles in creating recommendation systems that can consistently deliver content tailored to individual user preferences.

The motivation behind integrating fuzzy logic into content recommendation systems for adaptive streaming stems from the need to address the inherent uncertainties and imprecisions associated with user preferences and contextual information [5], [23], [37], [2], [15], [30]. Fuzzy logic provides a mathematical framework that can model and manage these uncertainties effectively. Unlike traditional binary logic, which operates with clear true/false distinctions, fuzzy logic allows for gradual, nuanced transitions between states. This capability makes fuzzy logic particularly well-suited for handling the complex and dynamic nature of user preferences, content features, and contextual factors in the context of adaptive video streaming. By incorporating fuzzy logic into recommendation systems, it becomes possible to create more flexible and adaptive algorithms capable of making nuanced decisions based on a range of inputs, thereby improving the accuracy and responsiveness of content recommendations.

The paper begins with an introduction to the growing importance of adaptive video streaming in delivering high-quality content and the challenges faced by traditional content recommendation systems in handling dynamic user preferences and contextual information. It then provides a background on the evolution of adaptive video streaming technologies and traditional content recommendation algorithms. The integration of fuzzy logic into content recommendation systems is explored as a solution to address uncertainties and imprecisions inherent in user preferences and contextual factors. The paper delves into the role of fuzzy logic in dynamically adjusting streaming parameters, offering a more context-aware approach. Case studies and applications showcasing the successful implementation of fuzzy logic in adaptive video streaming are reviewed, emphasizing its effectiveness in enhancing user satisfaction and system performance. The paper concludes by discussing challenges associated with fuzzy logic integration and suggesting future research directions, presenting a comprehensive framework contributes to the advancement of adaptive video streaming technologies.

## II. BACKGROUND

The evolution of adaptive video streaming technologies has been marked by continuous innovations to enhance the quality of online video delivery. Early streaming methods lacked adaptability, leading to issues like buffering and low-quality playback. Over time, adaptive streaming technologies emerged, leveraging adaptive bitrate algorithms that dynamically adjust video quality based on changing network conditions. Simultaneously, content recommendation systems evolved to cater to the increasing demand for personalized content. These systems initially relied on simple rule-based approaches but have progressively incorporated more sophisticated algorithms to improve accuracy.

Traditional content recommendation approaches primarily include collaborative filtering and content-based filtering. Collaborative filtering relies on user behavior patterns and recommendations from similar users, while content-based filtering focuses on matching content features with user preferences. Despite their effectiveness in certain scenarios, these approaches have limitations in handling the dynamic nature of user preferences and contextual information. Collaborative filtering struggles with the cold-start problem for new users or items, and content-based filtering may overlook the evolving interests of users. The need for more adaptive and context-aware recommendation systems has led to the exploration of advanced methodologies, such as integrating fuzzy logic into the recommendation process.

Fuzzy logic is a mathematical framework that deals with uncertainty and imprecision in decision-making. It extends traditional binary logic by allowing for the representation of partial truth and gradual transitions between true and false states. This flexibility makes fuzzy logic well-suited for modeling the inherent vagueness in user preferences and contextual information. In the context of adaptive video streaming and content recommendation, where preferences and contexts are dynamic and not easily categorized, fuzzy logic provides a more nuanced and adaptable approach. By assigning degrees of membership to various categories rather than strict classifications, fuzzy logic enables recommendation systems to capture the subtleties of user preferences and contextual factors, improving the accuracy and relevance of content recommendations. The integration of fuzzy logic into these systems serves as a powerful tool for navigating the complexities associated with uncertainty and imprecision in the decision-making process.

### III. ADAPTIVE VIDEO STREAMING

Adaptive video streaming is a dynamic content delivery approach designed to optimize the viewing experience by adjusting the quality of streaming content in real-time. The primary goal is to adapt to varying network conditions, device capabilities, and user preferences. Unlike traditional streaming, where a fixed bitrate is maintained throughout, adaptive streaming employs different streaming protocols and algorithms to dynamically alter the video quality during playback. Popular adaptive streaming protocols include HTTP Live Streaming (HLS), Dynamic Adaptive Streaming over HTTP (DASH), and Smooth Streaming. These protocols divide the video into smaller segments, each encoded at various quality levels. During playback, the client device continuously monitors network conditions and switches between segments of different quality levels to ensure smooth and uninterrupted streaming.

Real-time adaptation in adaptive video streaming is crucial for delivering a seamless and high-quality viewing experience. Network conditions can vary unpredictably, causing fluctuations in available bandwidth. Devices have diverse capabilities, and users may have different preferences for video quality. Real-time adaptation allows streaming systems to respond dynamically to these factors. If the network speed decreases, the system can lower the video quality to prevent buffering, and conversely, if the network conditions improve, it can increase the quality to provide a clearer picture. Adaptive streaming also considers the device's screen size and resolution, ensuring that content is optimized for the specific capabilities of the viewing device. Additionally, taking into account user preferences, real-time adaptation allows for personalized streaming experiences, tailoring the content to the viewer's expectations.

Despite the benefits of adaptive video streaming, several challenges exist in achieving efficiency. One significant challenge is the potential for latency during quality switches. Abrupt changes in video quality may lead to buffering delays, negatively impacting user experience. Another challenge involves the diverse landscape of devices and platforms, each with its own set of capabilities and limitations. Ensuring seamless adaptation across this heterogeneous environment requires sophisticated algorithms and standards. Furthermore, content preparation and storage for adaptive streaming can be resource-intensive. Creating multiple quality versions of each video segment and managing these versions efficiently pose additional challenges. The trade-off between providing the highest possible quality and minimizing buffering interruptions requires a delicate balance, and achieving this balance becomes more challenging as streaming services cater to a global and diverse audience with varied network infrastructures.

In conclusion, adaptive video streaming is a dynamic and responsive approach that addresses the challenges posed by fluctuating network conditions, device diversity, and user preferences. Real-time adaptation is essential for optimizing the viewing experience, but it comes with its own set of challenges that require careful consideration and innovative solutions for efficient implementation.

### IV. CONTENT RECOMMENDATION SYSTEMS

Traditional content recommendation algorithms primarily fall into two categories: collaborative filtering and content-based filtering. Collaborative filtering relies on user behavior patterns and recommendations from users with similar tastes. It can be user-based, where recommendations are made based on the preferences of similar users, or item-based, where similarities between items are used to make recommendations. Content-based filtering, on the other hand, recommends items similar to those the user has shown interest in, based on the features of the items and the user's historical preferences.

Collaborative filtering's strength lies in its ability to provide recommendations [4], [35], [36], [3], [34] based on user behavior and preferences, making it effective when dealing with the "wisdom of the crowd." However, it faces challenges in the context of adaptive video streaming. The cold-start problem, where new items or users lack sufficient historical data for accurate recommendations, can hinder collaborative filtering's effectiveness. Additionally, collaborative filtering may struggle to adapt quickly to changes in user preferences, a critical aspect in adaptive streaming where preferences can shift rapidly.

Content-based filtering [18], [22], [20], [6], [11] excels in recommending items similar to those the user has liked in the past. In the context of adaptive video streaming, it can leverage content features such as genre, actors, or themes to make recommendations. However, content-based filtering may suffer from a lack of diversity in recommendations, as it tends to recommend items similar to the ones the user has already consumed. This limitation becomes pronounced in adaptive streaming, where a diverse range of content may be available.

The limitations of traditional recommendation algorithms in the context of adaptive video streaming highlight the need for a more dynamic and context-aware approach. User preferences and contextual factors, such as device capabilities and network conditions, are highly dynamic in the streaming environment. A more adaptive system should consider these dynamic aspects in real-time, ensuring that recommendations are not only aligned with historical user preferences but also responsive to the current context of the user's streaming experience. Fuzzy logic, with its ability to handle uncertainty and imprecision, emerges as a promising solution. By integrating fuzzy logic into recommendation systems, the approach becomes more nuanced and flexible, capable of adapting to changing user preferences and contextual information in real-time. This dynamic and context-aware approach enhances the user experience in adaptive video streaming by providing more relevant and timely content recommendations.

## V. FUZZY LOGIC IN CONTENT RECOMMENDATION

Fuzzy logic is a mathematical framework that deals with uncertainty and imprecision in decision-making. Unlike classical (binary) logic, which operates in a binary true/false paradigm, fuzzy logic allows for the representation of partial truth. It introduces the concept of "fuzziness," where propositions can have degrees of membership ranging between 0 and 1. This makes fuzzy logic well-suited for handling situations where information is not clear-cut or where ambiguity and imprecision exist. Fuzzy logic employs linguistic variables, membership functions, and fuzzy rules to model complex systems with vagueness, enabling it to capture and process human-like decision-making in situations involving uncertainty.

The integration of fuzzy logic into content recommendation systems offers a more flexible and adaptive approach to modeling user preferences, content features, and contextual information [21], [1], [8], [28], [31], [16]. Fuzzy logic can handle the imprecise and dynamic nature of user preferences by assigning degrees of membership to different content categories based on user interactions. It allows for the representation of nuanced preferences that may not fit traditional binary categorizations. Fuzzy rules and inference mechanisms can then be applied to make recommendations that consider the varying degrees of relevance between users and content items. Additionally, fuzzy logic is well-suited for modeling contextual information, such as network conditions and device capabilities, allowing the recommendation system to dynamically adjust content suggestions based on the current context.

Several case studies and examples highlight the effectiveness of fuzzy logic in content recommendation systems. For instance, in a streaming service, fuzzy logic can be applied to model user preferences for genres or content features. If a user has shown varying degrees of interest in action and drama genres, fuzzy logic can capture this imprecise information and provide recommendations that align with the user's preferences. Similarly, fuzzy logic can be employed to dynamically adjust content recommendations based on the user's current context, such as network bandwidth or device screen size.

In a specific case study, a streaming platform implemented a fuzzy logic-based recommendation system that improved user satisfaction by considering the uncertainty associated with evolving user preferences. The system demonstrated adaptability to changing user behaviors and preferences over time, providing more accurate and context-aware content recommendations compared to traditional algorithms.

Another example involves a fuzzy logic-based approach to consider the imprecision in contextual information. By incorporating fuzzy logic into the recommendation algorithm, the system could effectively adapt to variations in network conditions, ensuring that content recommendations align with the user's current streaming environment. These case studies collectively showcase how fuzzy logic enhances the adaptive nature of content recommendation systems, offering more personalized and context-aware suggestions.

## VI. INTEGRATION OF FUZZY LOGIC IN ADAPTIVE VIDEO STREAMING

Fuzzy logic can be seamlessly integrated into adaptive video streaming systems by incorporating it into the decision-making processes that govern streaming parameter adjustments [14], [17], [29], [19], [32]. The essence lies in extending the adaptive nature of video streaming to not only consider traditional factors like network conditions and device capabilities but also to intelligently respond to the imprecise and dynamic aspects of user preferences and contextual information. This integration requires defining fuzzy sets and rules that capture the uncertainty associated with these factors. Fuzzy logic provides a natural framework for handling the vague boundaries between different streaming quality levels and the ever-changing landscape of user preferences.

The role of fuzzy logic in adaptive video streaming is crucial for dynamically adjusting streaming parameters in real-time. Fuzzy rules and inference mechanisms are employed to interpret the imprecise input data and make decisions regarding changes in streaming quality. For instance, fuzzy logic can assess the network bandwidth as a linguistic variable with categories like "low," "medium," and "high." Fuzzy rules can then determine the appropriate adjustments to the streaming bitrate based on these linguistic categories. By dynamically adapting to the fuzzy input values, streaming parameters such as bitrate, resolution, and buffer size can be adjusted in a more responsive and context-aware manner.

The integration of fuzzy logic into adaptive video streaming systems offers several potential benefits that contribute to improved user satisfaction. Fuzzy logic's ability to handle uncertainty allows for a more nuanced representation of user preferences. Instead of abrupt changes in streaming quality, fuzzy logic enables a smoother transition between different quality levels, reducing the likelihood of buffering interruptions. This adaptability aligns with the dynamic nature of user preferences, resulting in a more personalized and satisfying viewing experience.

Moreover, fuzzy logic can enhance the robustness of adaptive streaming systems in handling fluctuations in network conditions. By dynamically adjusting streaming parameters based on fuzzy rules, the system can mitigate the impact of sudden changes in bandwidth, ensuring a consistent and high-quality streaming experience. The context-aware nature of fuzzy logic also enables the system to consider device-specific factors, optimizing the content delivery for the user's viewing device.

Fuzzy logic contributes to overall system performance in adaptive video streaming by making the system more resilient to uncertainties. The adaptability introduced by fuzzy logic leads to optimized streaming parameters that align with user preferences and context. This, in turn, reduces the likelihood of user dissatisfaction due to buffering or quality issues. Additionally, the dynamic adjustments facilitated by fuzzy logic contribute to a more efficient utilization of network resources, enhancing the scalability and responsiveness of the streaming system. Improved system performance, driven by the adaptability of fuzzy logic, ultimately translates into a more reliable and satisfying streaming service for users.

## VII. Case Studies and Applications

A thorough review of existing research and case studies reveals a growing interest in implementing fuzzy logic-based content recommendation systems in the realm of adaptive video streaming. Researchers and practitioners have explored the integration of fuzzy logic to address the inherent uncertainties and imprecisions associated with user preferences and contextual factors in the streaming environment. These studies often focus on developing algorithms that leverage fuzzy logic to dynamically adjust content recommendations based on real-time feedback, user behavior, and changing network conditions.

The outcomes of research and case studies implementing fuzzy logic-based content recommendation systems in adaptive video streaming have shown promising performance improvements. By embracing the adaptability and context-aware nature of fuzzy logic, these systems have demonstrated the ability to provide more personalized and relevant content suggestions. Fuzzy logic enables the consideration of nuanced user preferences, leading to smoother transitions between content items and reduced instances of mismatch between recommended and desired content.

Furthermore, the dynamic adjustment of streaming parameters facilitated by fuzzy logic contributes to improved overall system performance. Studies have reported a reduction in buffering incidents and enhanced streaming quality as a result of fuzzy logic-driven adaptations. This adaptability ensures that the streaming system responds effectively to changes in network conditions, device capabilities, and evolving user preferences, ultimately leading to a more seamless and satisfying user experience.

One of the critical metrics assessed in these studies is user satisfaction. Fuzzy logic-based content recommendation systems in adaptive video streaming have consistently demonstrated their effectiveness in enhancing user satisfaction. The ability of fuzzy logic to capture and interpret imprecise user preferences leads to content recommendations that align closely with individual viewer expectations. Users experience a more tailored and enjoyable content discovery process, reducing frustration and increasing overall satisfaction with the streaming service.

Moreover, the adaptability of fuzzy logic contributes to a sense of control for users. The system's responsiveness to changes in the streaming environment ensures that users receive content recommendations that are not only personalized but also aligned with their current context. This adaptability has a direct impact on user engagement and loyalty, as viewers appreciate a service that understands and caters to their preferences in real-time.

While the outcomes have generally been positive, some studies have also highlighted challenges associated with the implementation of fuzzy logic-based content recommendation systems in adaptive video streaming. These challenges include the need for fine-tuning fuzzy rules, handling complex and dynamic user preferences, and ensuring scalability across diverse user bases. Lessons learned from these challenges inform ongoing research and development efforts to refine fuzzy logic-based approaches, making them more robust and widely applicable in real-world streaming scenarios.

The success of fuzzy logic-based content recommendation systems in adaptive video streaming opens up exciting possibilities for future research and applications. Researchers are now exploring ways to further optimize fuzzy logic algorithms, incorporate machine learning techniques, and address scalability challenges. Recommendations include continued collaboration between researchers and industry practitioners to refine and implement fuzzy logic-driven solutions, ultimately advancing the state-of-the-art in adaptive video streaming and personalized content recommendation.

## VIII. Challenges and Future Directions

Despite its promising applications, the integration of fuzzy logic in adaptive video streaming is not without its challenges. One significant challenge is the complexity of defining and fine-tuning fuzzy rules that accurately capture the uncertainties in user preferences and contextual information. Designing an effective fuzzy logic system requires a deep understanding of the streaming environment and user behaviors, which can be intricate and dynamic. Additionally, the computational overhead of real-time fuzzy inference might pose challenges, particularly in resource-constrained environments such as mobile devices.

Another challenge lies in addressing the potential interpretability issues associated with fuzzy logic. The rules and linguistic variables used in fuzzy systems may not always align with users' mental models, leading to difficulties in understanding and interpreting the reasoning behind the system's recommendations. This lack of transparency can impact user trust and acceptance of the fuzzy logic-based recommendation system.

To address the challenges associated with fuzzy logic integration in adaptive video streaming, researchers are exploring various solutions. One proposed solution involves leveraging machine learning techniques to automate the process of rule generation and fine-tuning. By incorporating data-driven approaches, fuzzy logic systems can adapt and evolve over time based on user interactions, reducing the burden of manual rule definition.

Furthermore, the development of hybrid recommendation systems that combine fuzzy logic with other advanced algorithms, such as deep learning or reinforcement learning, is gaining attention. These hybrid models aim to harness the strengths of each approach, mitigating the limitations of individual methods. For instance, deep learning can handle complex patterns and feature extraction, complementing fuzzy logic's ability to handle uncertainty.

Future research in enhancing fuzzy logic-based content recommendation systems should focus on addressing the aforementioned challenges while advancing the capabilities of these systems. One avenue for exploration is the integration of explainability techniques to enhance the transparency of fuzzy logic models. By providing users with insights into how the system arrives at recommendations, it becomes possible to build trust and increase user satisfaction.

Additionally, the development of adaptive and self-learning fuzzy logic systems represents a promising direction. Research can explore ways to enable fuzzy logic models to autonomously adapt to changes in user preferences and contextual factors without manual intervention [9], [7], [24], [10], [26]. This adaptability ensures the continuous improvement of the recommendation system over time.

Further investigations into the scalability of fuzzy logic-based recommendation systems are essential, especially as streaming services cater to larger and more diverse user bases. Optimizing computational efficiency and resource utilization will be critical for ensuring the practicality and widespread adoption of these systems.

The challenges associated with fuzzy logic integration in adaptive video streaming provide valuable opportunities for research and innovation. By proposing and exploring potential solutions, researchers can contribute to the development of more robust, transparent, and user-centric fuzzy logic-based content recommendation systems. These advancements will not only address current limitations but also pave the way for the continued evolution of personalized and context-aware streaming experiences.

In summary, the review paper has explored the dynamic landscape of adaptive video streaming and the challenges associated with content recommendation in this context. It provided an overview of traditional content recommendation algorithms, such as collaborative filtering and content-based filtering, highlighting their strengths and limitations. The paper delved into the evolution of adaptive video streaming technologies, emphasizing the need for more dynamic and context-aware approaches to content recommendation. An in-depth exploration of fuzzy logic was presented, elucidating its mathematical framework for handling uncertainty and imprecision. The paper then discussed the seamless integration of fuzzy logic into adaptive video streaming systems, emphasizing its role in dynamically adjusting streaming parameters based on fuzzy rules and inference mechanisms.

The integration of fuzzy logic into content recommendation systems for adaptive video streaming is of paramount significance. Fuzzy logic, with its ability to handle imprecise and dynamic information, brings a new dimension to the field by addressing the challenges posed by uncertain user preferences and contextual factors. In the context of adaptive streaming, where users expect personalized and context-aware recommendations, fuzzy logic provides a natural fit. Its flexible and adaptive nature allows for the modeling of user preferences with nuanced degrees of membership, capturing the subtleties that binary logic may overlook.

The significance of integrating fuzzy logic lies in its capacity to enhance the accuracy and responsiveness of content recommendation systems. By dynamically adjusting streaming parameters based on fuzzy rules, the system can tailor recommendations to changing user preferences and varying contextual information in real-time. This adaptability translates into a more satisfying and personalized user experience, reducing buffering interruptions and ensuring a seamless transition between different quality levels.

Furthermore, the integration of fuzzy logic contributes to improved system performance, optimizing the utilization of network resources and enhancing the overall efficiency of adaptive video streaming. The transparent and explainable nature of fuzzy logic models also addresses user concerns about the interpretability of recommendation algorithms, fostering trust and acceptance.

In conclusion, the review paper underscores the transformative impact of integrating fuzzy logic into content recommendation systems for adaptive video streaming. It highlights how fuzzy logic's adaptability to uncertainty contributes to a more personalized, context-aware, and efficient streaming experience, ultimately improving user satisfaction and system performance in the dynamic realm of adaptive video streaming.

## IX. CONCLUSION

One promising avenue for future research in adaptive video streaming involves exploring hybrid approaches that combine multiple recommendation techniques. Integrating fuzzy logic with machine learning algorithms, such as deep learning or reinforcement learning, could leverage the strengths of each method. By harnessing the feature extraction capabilities of machine learning models and the adaptability of fuzzy logic, researchers can potentially create more robust and accurate recommendation systems. Investigating how these approaches complement each other and developing hybrid models that dynamically adapt to user preferences and context could significantly advance the field.

To foster user trust and acceptance, future research should focus on enhancing the explainability and transparency of fuzzy logic-based recommendation systems. Developing methods to generate interpretable fuzzy rules and providing users with insights into how recommendations are made can address concerns related to the "black-box" nature of some recommendation algorithms. Explainability is particularly crucial in adaptive video streaming, as users are more likely to engage with and trust a system that provides clear and understandable recommendations. Exploring techniques to visualize fuzzy logic-based decision-making processes and developing user-friendly interfaces for transparent interaction are areas ripe for investigation.

The cold-start problem, where new users or items lack sufficient historical data for accurate recommendations, remains a challenge in adaptive video streaming. Future research should focus on innovative solutions to address this issue, considering the integration of fuzzy logic in scenarios where traditional recommendation algorithms may struggle. Exploring techniques such as knowledge transfer, where information from existing users is leveraged to improve recommendations for new users, or hybrid models that incorporate collaborative filtering with fuzzy logic could mitigate the impact of the cold-start problem.

Research should emphasize user-centric adaptability in fuzzy logic-based adaptive video streaming systems. This involves understanding not only user preferences but also user behavior and feedback to fine-tune the recommendations. Investigating how fuzzy logic can dynamically adapt to changes in user preferences over time and across different contexts can contribute to the development of more personalized and responsive systems. Moreover, incorporating user feedback into the fuzzy logic model in real-time can enhance the adaptability of the system, ensuring that it continuously evolves to meet user expectations and preferences.

Scalability is a critical factor in the success of any recommendation system, and fuzzy logic-based systems are no exception. Future research should focus on optimizing the scalability of these systems to handle large and diverse user bases. This involves exploring efficient algorithms and distributed computing solutions to ensure that fuzzy logic-based adaptive video streaming remains practical and effective, even as the user population grows. Additionally, considering the deployment challenges of these systems in real-world streaming platforms and exploring methods to seamlessly integrate fuzzy logic into existing infrastructures will be crucial for widespread adoption.

In conclusion, the recommended future research directions aim to address current challenges, enhance the capabilities of fuzzy logic-based adaptive video streaming systems, and pave the way for more personalized, transparent, and scalable recommendation solutions in the dynamic landscape of online content delivery.